\documentclass{bmcart}

\usepackage{graphicx}
\usepackage{longtable}
\usepackage[utf8]{inputenc} 
\usepackage{nomencl}
\makenomenclature

\startlocaldefs
\endlocaldefs

\begin{document}

\begin{frontmatter}

\begin{fmbox}
\dochead{Research}


\title{Models towards Risk Behavior Prediction and Analysis: A Netherlands Case study}


\author[
   addressref={aff1},                   
   corref={aff1},                       
   email={seun.adekunle@maastrichtuniversity.nl}   
]{\inits{OH}\fnm{Onaopepo} \snm{Adekunle}}
\author[
   addressref={aff2},
   email={a.riedl@maastrichtuniversity.nl}
]{\inits{A}\fnm{Arno} \snm{Riedl}}
\author[
   addressref={aff1},
   email={michel.dumontier@maastrichtuniversity.nl}
]{\inits{M}\fnm{Michel} \snm{Dumontier}}


\address[id=aff1]{
  \orgname{Netspar \& Institute of Data Science, Maastricht University}, 
  \city{Maastricht},                              
  \cny{NL}                                    
}
\address[id=aff2]{%
  \orgname{CESifo, IZA, Netspar \& Department of Microeconomics and Public Economics, School of Business and Economics, Maastricht University},
  \city{Maastricht},
  \cny{NL}
}
\address[id=aff3]{%
  \orgname{Centraal Bureau voor de Statistiek},
  \city{Meerssen},
  \cny{NL}
}


\begin{artnotes}
\end{artnotes}

\end{fmbox}


\begin{abstractbox}

\begin{abstract} 
\parttitle{Background} 
In many countries financial service providers have to elicit their customers risk preferences, when offering products and services. For instance, in the Netherlands pension funds will be legally obliged to factor in their clients risk preferences when devising their investment strategies. Risk preferences are also an important concept for investment behavior, consumer credit, insurance, pension plans, tax behavior, and becoming a victim of credit card fraud. Therefore, assessing and measuring the risk preferences of individuals is critical for the analysis of individuals' behavior and policy prescriptions.

\parttitle{Problem} 
In the psychology and economics a number of methods to elicit risk preferences have been developed using hypothetical scenarios and economic experiments. These methods of eliciting individual risk preferences are usually applied to small samples because they are expensive and the implementation can be complex. are organizationally complex to perform on large representative samples. However, for practical application often the risk preferences of large cohorts need to be measured. Data science methods exploiting existing register data on individual characteristics could be used to predict risk preferences of large cohorts and in that way mitigate the risk preferences elicitation challenge.

\parttitle{Approach} 
A large number of supervised learning models ranging from linear regression to support vector machines are used to predict risk preference measures using socio-economic register data such as age, gender, migration background and other demographic variables in combination with data on income, wealth, pension fund contributions, and other financial data. The employed machine learning models cover a range of assumptions and properties as well as a diverse set of regression metrics. The optimum model is selected using the metrics and interpretabilty of the model. The optimal models  are lasso regression and gradient boosting machines with mean average percentage error of about 30\%. This can be loosely interpreted as approximately 70\% accuracy for prediction of the individual risk preference.

\parttitle{Significance} 
The results are informative regarding which of the above factors (features) are most strongly correlated with individuals' risk preferences. This is important as it helps determining risk attitudes without actually measuring them. It should be noted that with the current accuracy the tested models are not ready for deployment for applications that require high accuracy. However, the results do indicate which models should be used in situations that do not require the most accurate predictions such as augmentation data for pensions' recommendation.
\end{abstract}




\end{abstractbox}
%

\end{frontmatter}



\newpage

\section{Introduction}
Over the last century, human life expectancy has increased globally. This positive trend also leads to challenges as to guarantee social well-being for the elderly it needs to be combined with a satisfactory standard of living during retirement.
For a long time this could be sustained with high economic growth and high returns on financial markets. However, currently high consumption levels at old age may not be guaranteed anymore because of limited economic performance and increased uncertainties due to climate change and political and military conflicts. This poses a challenge for politics as well as the pension industry and insurance sector. Specifically, companies offering life insurance products and pension plans and governments providing retirement income are facing challenges due to the increased life expectancy in combination with uncertain future economic and finance performance \cite{levantesi2021a}. To handle these challenges new policies are required that provide more tailor-made products. This, however, requires better knowledge of the individual preferences. As a case at hand, in the new pension agreement, the Netherlands require from the pension funds that they measure the individual risk preferences of their clients and base the investment strategies on age cohort specific risk preferences \cite{koolmees2019a}. This highlights that good estimates of individual risk preferences are important and will become more so in the future. 

Risk is ubiquitous in decision-making and individuals' risk preferences are considered to be one of the most important determinants regarding the  extent to which people are willing to take risk. Risk preferences are important as they affect behavior in may domains such as investment, consumer credit, insurance and pension plans, and tax compliance, to name just a few. In addition, assessing and measuring risk preferences of individuals is critical for economic analysis and policy prescriptions. In prominent theories of decision-making under risk and uncertainty such as subjective expected utility theory \cite{savage1954a} and prospect theory \cite{tversky1979a} risk preferences are an important parameter.

As an example, consider the case where a person is given a chance to purchase a lottery ticket with equal chances of winning either €10 or €0. A risk-neutral individual will be willing to pay up to €5—-the expected value of the lottery--in order to play the lottery . Individuals who are only willing to pay less than €5 are considered to be risk-averse, while those willing to pay more than €5 are considered to be risk-seeking.

One challenge in measuring risk preferences is that they differ largely across individuals. To account for that, economists and psychologists have developed a variety of survey and experimental methods to elicit and assess individual risk attitudes. Bokern and co-authors \cite{bokern2021a} describe the various methods to elicit risk preferences and conclude that there is not a unique optimal one. In reviewing the evidence they also find that the predictive power regarding actual behavior of elicited risk preferences using these methods appears to be limited. 

This paper uses Machine Learning (ML) techniques to explore the extent to which elicited risk preferences are correlated with individual background characteristics exploiting readily available register data. Two risk preferences elicitation methods will be scrutinized: self reported measures of risk preferences using questionnaires and incentivized measures with lottery choices used in experimental economics research. The paper also discusses the practical applicability of risk preferences predicted with ML methods.

ML algorithms are rooted in computer science and statistics and focus on minimizing the out-of-sample prediction error \cite{athey2017a, mullainathan2017a}. They are gaining popularity among economists and other social scientists, because they provide them with an alternative tool which is useful especially for predictive tasks \cite{varian2014a}. In particular, a relatively recent stream of research has focused on the so-called “prediction policy problems” \cite{athey2017b, chalfin2016a, kleinberg2016a}, which has been introduced by Kleinberg and co-authors \cite{kleinberg2015a}.

To understand the wide inter-individual variety of risk preferences, large scale questionnaire or experimental studies would be needed, preferably repeatedly. However, their implementation is organizationally complex and resource intensive. Therefore, the out of sample prediction of elicited risk preferences from available register data on individual characteristics investigated here is potentially very useful. An important additional insight delivered by the presented study is that it looks at how ML methods can be utilized to gain further insight on the set of variables that correlate with individual risk preferences. Specifically, ML can be leveraged to make quick estimates for a large population based on measures derived from a small sample. In principle, the derived estimates are deployable in larger settings such as pension plans and savings recommendations for a whole population.

\section{Literature review: ML in economics and finance}
There has been an increasing interest in the usage of ML methods in finance and economics because of its capacity to handle big data. ML applied to economic problems can be traced back as early as 1974 \cite{lee1974a}, although there it is merely mentioned in the abstract. In 1988, White \cite{white1988a}, published a paper that involved a Neural Networks (NN) application to forecast the IBM daily stock returns. Since then, the application of ML in economics steadily increased. Initially, it was applied for forecasting financial time series where long datasets are widely available. ML systems from that era required extensive data sets for efficient training, that did not exist in other areas of economics. Training was also very time-consuming due to the comparatively low processing power of the computers of that period. Nowadays, the use of many new ML architectures that do not require unreasonably long data sets, is an interesting and very promising avenue in forecasting tasks in economics and finance. This holds also for applications in macroeconomics or microeconomics where data sets are often limited in size. Recent ML applications in, for example, business cycles research and recession forecasting appear to be very successful when compared to more traditional empirical approaches\cite{Athey2019MachineLM, Hall2018MachineLA, Kreiner2019CanML}. Also, new methodologies merging and combining econometrics with ML (e.g. Garch–SVM) are on the rise\cite{garch_combined}. 

Recent research used ML to predict annual wages in a Population Survey utilizing a small subset of variables and an expanded set of variables \cite{ghei2020a}. The authors employed XGBoost (Extreme Gradient Boosting), LightGBM (Light Gradient Boosting Machine), and Deep Learning together with stacking techniques to achieve better prediction accuracy. In that way they were able to achieve a prediction improvement over the baseline Mincer regression even with a subset of their variables.

Another study explored the factors related to individuals’ risk perceptions associated with COVID-19 as well as their general self-assessed risk preferences \cite{kassas2021a}. They find that both risk perceptions and preferences are significantly correlated with mitigation behaviors practiced during the pandemic and that risk perceptions are correlated with a larger number of mitigation practices. In a different domain, researchers used a lottery-choice mechanism to measure farmers' preferences over money-denominated risks \cite{hellerstein2013a}. They looked at whether these choice data predict farming decisions for a sample of 68 farmers. They concluded that their risk preference measure has explanatory power for the prediction of farming decisions in the study but it is still unproven whether it is indicative of general farmer behavior.

Another vein of research assumes that the decision to take risk causes decision makers to integrate gains, losses, and outcome probabilities which is represented by conflict resolution. The study by Stillman and co-authors \cite{stillman2020a} employs the use of dynamic assessments of choices such as mouse tracking to index conflict. They propose that mouse-tracking metrics of conflict sensitively detect differences in the subjective value of risky versus certain options thereby predicting the participants’ risk preferences. A similar study by Aimone and co-authors \cite{aimone2016a} tracks the eye movement of participants rather than mouse clicks to predict risk preference of participants.

To the best of our knowledge, we are the first using ML techniques, to explore the possibility of relating risk preferences that are elicted with survey questions and incentivized experiments, respectively, to individual characteristics gained from register data.

\section{Methodology}
\subsection*{Data}
The data collected for this analysis consists of a two-wave online survey in May and June of 2020, carried out with the aid of research agency Flycatcher. Statistics Netherlands (CBS) selected a stratified random sample of 18,000 Dutch employees and 18,000 self-employed who were invited to participate in the study. In total, 4,282 Dutch residents completed both waves. Data from the survey is enriched with demographic and socioeconomic variables from register data of CBS. 
Using the CBS data, we classify 2,224 (52\%) as employed, 1,480 (35\%) as self-employed, and 190 (4\%) as other (e.g., student, retiree, unemployed). We exclude participants classified as other because they are neither employed nor self-employed. 

The two waves of the survey included a different set of incentivized elicitation tasks. The incentivized behavioral measure for risk preferences consisted of five different multiple price lists (MPLs) in the tradition of Holt and Laury behavioral economics paper \cite{holt2002risk}. An MPL is a list of binary decision situations. In the case of risk preferences, participants are asked to choose between a safer and riskier lottery in each decision situation. The list is designed such that either the safer or the riskier lottery becomes more attractive when moving down the list. The point where the participant switches to the option that becomes more attractive provides an indication of the risk preference. In this study, participants made ten choices in each MPL. We take the average number of safe lottery choices over all five MPLs as a measure for risk preference. The self-reported survey questions are based on the work by Dohmen and co-authors \cite{dohmen2011}. Participants identified themselves as being more or less willing to take risk on an 11-point Likert-scale ranging from “not at all willing to take risks” (0) to “very willing to take risk” (10) in a general domain and several specific life domains. One in five participants, among those who completed both waves, was randomly selected to be paid out based on their decisions in one randomly selected task. In addition, one iPad was raffled off among the participants who completed both waves. Earnings ranged from €0 up to €186 depending on the task. The average earning among the participants selected for payment was €78. Participants were fully informed about the procedures in advance. The appendix provides more details on the survey.

The survey data has been used to estimate the risk preference of all participants which is the outcome to be estimated using supervised machine learning techniques. The register data from CBS consists of 65 features which serve as the predictors and the square of the age feature as a derived feature, ranging from demographic features such as age, migration background, gender etc. to financial features like income, wealth, debt, pension fund etc. The age squared is included due to some evidence in the literature of its relationship with risk preference \cite{falk2018a}.  Features are classified as either personal features, like age, sex, income, pension etc. or household features, like household income, breadwinner.

The risk preference serves as the target of prediction in a supervised learning task. The is estimated by finding the average of selected options from a multiple price list methodology of estimating risk preference \cite{bokern2021a} and Likert scale survey results of participants stating how much they are willing to take risk. 

\subsection{Data Processing}
The total 66 predictor features consist of 38 categorical variables and 28 numerical variables with average missing value percentage of 12\%. After inspection of the data, all variables with less than 10 value counts and string variables were considered categorical.

\subsubsection{Missing Data Imputation}
Various imputation methods have been tested on the data ranging from simple averages to more complex iterative imputation techniques. The methods that yielded the optimal results upon model tuning were M-estimate categorical encoding and LightGBM iterative imputation for numerical features.

\subsubsection{M-estimate Encoding}
This is a type of target encoding where missing values are replaced with the mean of the target variable but compensates for over-fitting using smoothing. It is also often called additive smoothing \cite{micci-barreca2001a}. 
\[
M-estimate_{encoding} = \frac{\bigl[ count(category) * mean(category) \bigr] + M * mean(target)}{count(category) + M}
\]
where $category$ is a unique categorical feature value, $target$ is the outcome feature value and $M$ is the smoothing parameter.

\subsubsection{Iterative imputation}
Iterative imputation refers to a process where each missing feature is modeled as a function of the other features, e.g. a regression problem where missing values are predicted. Each feature is imputed sequentially, allowing prior imputed values to be used as part of a model in predicting subsequent features. It is iterative because this process is repeated multiple times, allowing ever improved estimates of missing values to be calculated as missing values across all features are estimated. The modeling function for our case is the light gradient boosting machine \cite{ke2017a}.

\subsection{Feature Selection}
Feature selection has been done through recursive feature elimination \cite{guyon2003a}, in particular with cross validation (RFECV). In RFECV, a model is repeatedly trained multiple times and each time the least important feature is removed  from the model, determined by coefficient value or feature importance attribute of the model. This process is then repeated until the model performance becomes worse, at which point, only the features that are most important are left. A ten-fold cross validation is also applied on the dataset. 

\subsection{Model Training}
In a view towards assessment of different ML techniques’ ability to predict the risk preference measures of interest, we considered a slew of regression models including ordinary least squares (OLS), ridge regression , lasso regression, least angle regression, bayesian ridge, elastic net, decision tree regression, random forest regression, extra trees regression, adaboost, gradient boosting, LightGBM, catboost, K-nearest neighbor, and huber regression. A dummy regressor that always outputs the mean risk preference measure is also trained as a baseline for the other methods. 
While it is not feasible to consider all possible ML algorithms, these cover parametric and non-parametric approaches, with varying abilities to account for non-linearity and dimensionality,  given the size of the dataset.Figure 1 provides an overview of the ML pipeline implemented for the workflow.

\begin{figure}[!htb]
    \centering
    \includegraphics[width=0.84\linewidth]{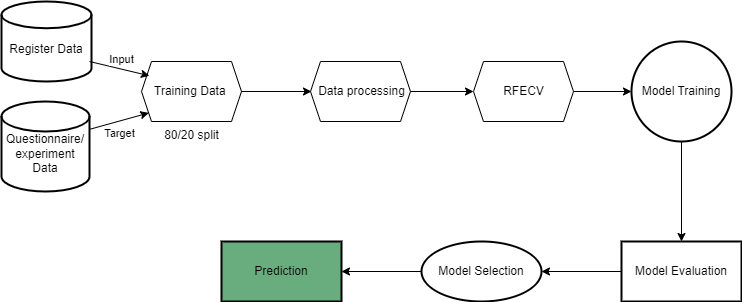}
    \caption{Overview of the workflow ML pipeline for prediction of risk preference.}\label{fig: figure1}
\end{figure}

Linear regression (OLS) fits a straight line on data points to minimize error between the line and the data points (Hastie, 2009) \cite{hastie2009a}. The relationship between the independent variable and the dependent variables is assumed to be linear. One of the major issues with OLS is that it is prone to overfitting, hence, ridge and lasso regression tries to alleviate the issue building upon OLS. Ridge regression (Hastie, 2009)\cite{hastie2009a} includes a constraint on the square value of the coefficient parameters to be estimated, while lasso (Hastie, 2009)\cite{hastie2009a} includes a constraint on the absolute value of the coefficient parameters to be estimated. Elastic net tries to combine both ridge regression and lasso to benefit from their advantages. Bayesian ridge regression is similar to the ridge regression, but it assumes normal likelihood and normal prior for the parameters.

Least-angle regression (LARS) is a regression algorithm for high-dimensional data that is similar to forward stepwise regression (Efron et al., 2004)\cite{efron2004a}. At each step, it finds the feature most correlated with the target. When there are multiple features having equal correlation, instead of continuing along the same feature, it proceeds in a direction equi-angular between the features.

A decision tree is applies the tree structure algorithm to analyze data. One major advantage is that some decision rules can be produced that are easy to understand by humans. A decision tree classifies an instance by sorting it through the tree to the appropriate leaf node, i.e. each leaf node represents a classification. Each node represents some attribute of the instance, and each branch corresponds to one of the possible values for this attribute. 

Random forest (RF) is a type of ensemble method that constructs a strong prediction model from many weak decision trees. RF builds a user predefined number of decision trees to form a forest, and then it takes the majority vote of these trees to output a single class prediction. For each tree, RF selects random samples with replacement and randomly generates a subset of features to decide each candidate node split; typically, the one with the highest Gini impurity or entropy. Extra trees regression is similar to the random forest, but it does not make use of sampling the data with replacement but uses the full data always with a random split criteria.

Gradient boosting (GBR) is an ensemble method that iteratively combines weak learners to build a strong learner. It uses decision trees as base learner and adaptively adds trees that minimizes a differentiable loss function. It is similar to adaptive boosting (adaBoost) which additionally adjusts the weights of misclassified samples at each iteration to focus on difficult samples and other boosting techniques such as:
Category boosting (CatBoost) that builds symmetric (balanced) trees, unlike LightGBM which is optimized for efficiency. In every step, leaves from the previous tree are split using the same condition. The feature-split pair that accounts for the lowest loss is selected and used for all the level’s nodes. This balanced tree architecture aids in efficient CPU implementation, decreases prediction time, makes swift model appliers, and controls overfitting as the structure serves as regularization. 

Category boosting is a type of machine learning algorithm that combines the power of boosting with categorical variables. It is particularly useful in situations where there are a large number of categorical variables, which are difficult to handle using traditional boosting techniques.

In CatBoost, the algorithm starts by creating a decision tree that uses one categorical variable as the root node. The algorithm then splits the data based on the categories of that variable and recursively creates sub-trees for each category. This process is repeated for each categorical variable until a stopping criterion is met.

The final model is a weighted sum of the individual decision trees, where the weights are determined during the boosting process. The output of the model is a probability distribution over the categories, which can be used to make predictions.

The category boosting algorithm can be formalized as follows:

Initialize the model by setting the initial weight of each sample to 1/N, where N is the number of samples.
For each boosting round:
a. Train a decision tree using the current weights and a randomly selected categorical variable as the root node.
b. Calculate the weighted error rate of the tree.
c. Calculate the weight of the tree based on its error rate.
d. Update the weights of the samples based on the classification error of the tree.
Output the final model, which is a weighted sum of the individual decision trees.
An example of CatBoost in action is in predicting the likelihood of a customer purchasing a product based on their demographic information. The demographic information may include categorical variables such as gender, age range, and location. By using CatBoost, the algorithm can handle these categorical variables effectively, and provide a reliable prediction of whether or not the customer is likely to purchase the product.

To assess the predictive ability and generalizability of the constructed models, we split the data into a training set and an independent test set with ratio 80\%/20\% by conducting a random sampling to make a representative distribution equal in both the training and test sets. We further split the training set, to create a validation set (20\% of the training set). A ten-fold cross validation on the training set including the validation set tunes each prediction model to find appropriate hyperparameters and to choose the optimal operating point. We then refit a model on the entire training set with the selected optimal hyperparameters and operating point to predict the test set.

\subsection{Evaluation}
Standard regression metrics such as mean absolute error (MAE), mean absolute percentage error (MAPE), mean square error (MSE), root mean square error (RMSE) and R-squared are used to evaluate the performance of the prediction models.

Mean Absolute Error (MAE), is a measure of prediction error, it is the average absolute difference between observed and predicted outcomes. MAE is less sensitive to outliers compared to RMSE. 
\[
MAE = \frac{1}{N}\sum_{i=1}^N |y_i - \hat{y}_i|,
\]
where $y$ is the observed outcome, $\hat{y}$ is the predicted outcome and $N$ is the sample size.

Mean Absolute Percentage Error, or MAPE provides an overall percentage indicating how accurate the prediction system is overall. It is MAE defined in percentage variation. MAPE is useful for understanding the average behavior of the prediction model and is especially helpful in communicating accuracy to people not familiar with other metrics. 
\[
MAPE = \frac{1}{N}\sum_{i=1}^N \frac{|y_i - \hat{y}_i|}{y_i} * 100
\]

Mean Squared Error (MSE) is similar to MAE but the prediction error is squared. It is defined as follows:
\[
MSE = \frac{1}{N}\sum_{i=1}^N (y_i - \hat{y}_i)^2
\]
That is, it is the average squared difference between the observed actual outcome values and the values predicted by the model.

Root Mean Squared Error (RMSE) is similar to the MSE as it is the square root of it, which is
\[
RMSE = \sqrt{\frac{1}{N}\sum_{i=1}^N (y_i - \hat{y}_i)^2}
\]

R-squared (R2), is the proportion of variation in the outcome that is explained by the predictor variables. In multiple regression models, R2 corresponds to the squared correlation between the observed outcome values and the predicted values by the model. The Higher the R-squared, the better the model.

\section{Results}
The prediction models were based upon two target variables. i) the average number of safe choices made by survey participants in a Multiple price list lottery (MPL) and ii) participants estimate their likelihood to take risk on a Likert scale directly asked to each individual. Figure 2 shows the distribution of the average number of safe choices risk preference measure and how it compares to a normal distribution. The variable has a distribution that is very similar to a normal distribution. This is reassuring as many models are assuming that the distribution is a bivariate or multivariate normal; e.g., Linear discriminant analysis, Gaussian Naive Bayes, Logistic Regression, Linear Regression, Bayesian models etc. 

\subsection{Model Interpretation}
\subsubsection{Average number of safe choices in MPL}
 The optimal model is selected based on two criteria i) the metrics evaluations and the interpretabilty of the model. Table ~\ref{Tab:t1} shows that the orthogonal  matching pursuit (OMP) has top mertic value performance out of all the metrics evaluated. It should be noted, however, that this employs a greedy algorithm that provides a hard constraint on the total number of non-zero coefficients in the model. This makes it more prone to over-training than the regularized OLS like ridge and lasso regression. Lasso regression comes third in terms of the metrics for fitting the data and has the added benefit of setting the coefficient of non-valuable features to zero; thus, providing an intuitive selection of relevant features and interpretation of results. It remains the optimal model for prediction of the average number of safe choices risk preference measure whilst preserving interpretability. The five top listed models in the table are the most appropriate ones for prediction because these have performed better than the dummy regressor. The full set of models trained and the metric performance is provided in the appendix section. 

\begin{figure}[!htb]
    \centering
    {\includegraphics[width=0.84\linewidth]{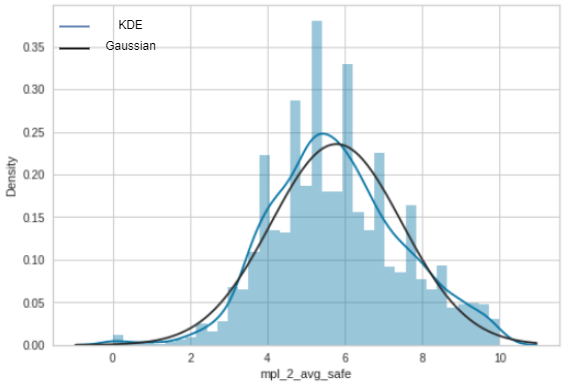} }

    \caption{The distribution of the average number of safe choices risk preference measure denoted as mpl-2-avg-safe compared to a Gaussian distribution.}\label{fig: figure2}
\end{figure}

The metric values of MSE, RSME and RMSLE are highly correlated as expected. However, the R-squared metrics in particular seems to be quite low which indicates that the variance explained by the models are not statistically significant, for example, R-squared of 0.001 in OMP which is the top model in terms of metrics; out of the maximum of 1. It should be noted that this does not necessarily mean that the models are bad, as it could be that the data are highly non-linear or may have many outliers.

It can be seen from the table that when using the MAPE, the OMP has the smallest error value, followed by elastic net and then lasso regression. MAPE has the advantage of easier interpretation for non-domain individuals, as it can be interpreted as the average error deviation of the predicted results to the gold standard. Hence, it can be interpreted as an accuracy metric in simplistic terms. When selecting the lasso regression model because of its aforementioned benefits, the MAPE is 0.2904 which is 29.04\% average error between the predicted result and the actual result. This can be loosely interpreted as 70.96\% accuracy of prediction results.

\begin{table}
\caption{Machine learning models for prediction of the average number of safe choices risk preference measure and the corresponding regression metric values.}
\centering
\setlength\tabcolsep{0pt}
\begin{tabular*}{\linewidth}{@{\extracolsep{\fill}} ||p{2.5cm} c c c c c c||} 
\hline \hline
\textbf{Model}                           & \textbf{MAE}    & \textbf{MSE}    & \textbf{RMSE}   & \textbf{R-Squared} & \textbf{RMSLE}  & \textbf{MAPE}   \\ 
\hline 
\textbf{Orthogonal Matching Pursuit}     & \textbf{1.3273} & \textbf{2.7990} & \textbf{1.6713} & \textbf{0.0140}    & \textbf{0.2786} & \textbf{0.2904} \\
\hline
\textbf{Elastic Net}                     & 1.3299          & 2.8111          & 1.6748          & 0.0097             & 0.2793          & 0.2909          \\
\hline
\textbf{Lasso Regression}                & 1.3306          & 2.8166          & 1.6765          & \textbf{0.0078}    & 0.2794          & \textbf{0.2908} \\
\hline
\textbf{Bayesian Ridge}                  & 1.3312          & 2.8198          & 1.6775          & 0.0067             & 0.2794          & 0.2908          \\
\hline
\textbf{Adaboost Regressor}              & 1.3418          & 2.8399          & 1.6833          & -0.0001            & 0.2813          & 0.2965          \\
\hline
\textbf{Dummy Regressor}                 & 1.3359          & 2.8422          & 1.6841          & -0.0011            & 0.2802          & 0.2916          \\
\hline
\end{tabular*}
\label{Tab:t1}
\end{table}
\begin{figure}[!htb]
    \centering
    {\includegraphics[width=0.84\linewidth]{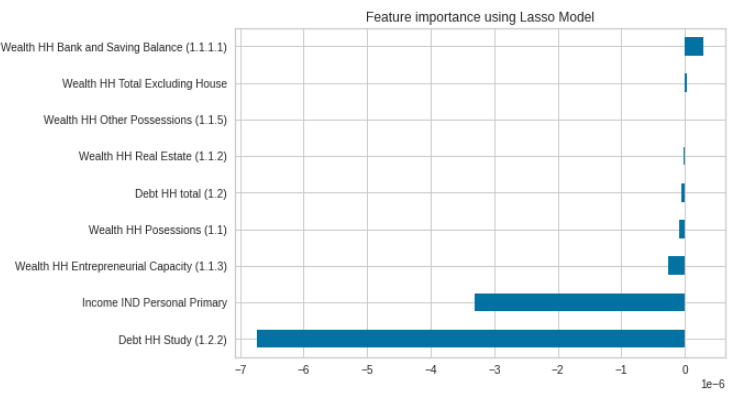} }

    \caption{Lasso regression coefficient of linear relationship to the average number of safe choices risk preference measure plot.}\label{fig: figure3}
\end{figure}
A more fine grained look at the results of the lasso model which is the optimal model for the dataset. Figure 3 shows for the lasso regression the feature importance for selected features. It can be seen that features related to finances such as wealth and income have the highest coefficient of linear relationship indicating their importance. Lasso eliminates 56 variables, setting their coefficient to zero, thereby indicating they do not contribute to the model result. The total household wealth excluding home ownership and the banking and saving balance household wealth are positively correlated to the average number of safe choices risk preference. While personal income, household debt, entrepreneurial capacity as a measure of household wealth and household wealth possessions are negatively correlated to the target feature.

\subsection{Survey Risk Preference}
\begin{figure}[!htb]
    \centering
    {\includegraphics[width=0.64\linewidth]{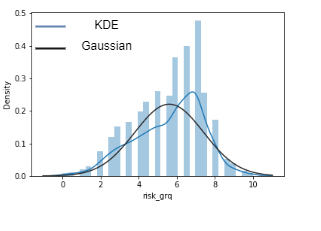} }

    \caption{The distribution and theoretical quantiles plot of the average number of safe choices risk preference measure denoted as risk-grq as compared to a normal distribution.}\label{fig: figure4}
\end{figure}

This section reports on the the outcomes using the same set of input features but with the target variable of the risk preference elicited from the survey participants. Figure ~\ref{fig: figure4} shows the distribution of the target variable and how it compares to a normal distribution. The distribution approximately fits a normal distribution. However, the fit is worse than for the target variable average number of safe choices. 

As can be seen from Table ~\ref{Tab:t2}, the top performing model in terms of metrics performance remains the OMP, except for the MAPE according to which the adaboost model performs best. The runner-up models in terms of metrics performance are gradient boosting machine models and bayesian ridge which also have approximately 36\% MAPE. The optimal model for this prediction task relative to metrics performance and interpretabilty is GBR with 36.56\% MAPE i.e. the average error between the predicted result and the actual result. Lasso regression results provides a basis of comparison between both target variables but in this case, its performance lies in the top ten with about 39\% MAPE. This is in our opinion still relatively acceptable as compared with the top performing models while outperforming the dummy regressor. The dummy regressor is outperformed by eleven models in this case which justifies the statement that there are more options for the prediction of the survey measured risk preference than the average number of safe choices risk preference measure.

\begin{table}
\caption{Machine learning models for prediction of self reported risk preference measure and the corresponding regression metric values.}
\centering
\setlength\tabcolsep{0pt}
\begin{tabular*}{\linewidth}{@{\extracolsep{\fill}} ||p{2.5cm} c c c c c c||} 
\hline \hline
\textbf{Model}                           & \textbf{MAE}    & \textbf{MSE}    & \textbf{RMSE}   & \textbf{R-Squared} & \textbf{RMSLE}  & \textbf{MAPE}   \\
\hline
\textbf{Orthogonal Matching Pursuit}     & \textbf{1.3846} & \textbf{2.9033} & \textbf{1.7037} & \textbf{0.1258}    & \textbf{0.3199} & 0.3658          \\
\hline
\textbf{Gradient Boosting Regressor}     & \textbf{1.3847}          & 2.9183          & 1.7081          & 0.1213             & \textbf{0.3201}          & \textbf{0.3656} \\
\hline
\textbf{Bayesian Ridge}                  & 1.3856          & 3.1209          & 1.7597          & 0.0617             & 0.3205          & 0.3647          \\
\hline
\textbf{Random Forest Regressor}         & 1.4044          & 2.9644          & 1.7215          & 0.1074             & 0.3212          & 0.3680          \\
\hline
\textbf{Catboost Regressor}              & 1.4234          & 3.0767          & 1.7537          & 0.0739             & 0.3248          & 0.3700          \\
\hline
\textbf{Light Gradient Boosting Machine} & 1.4259          & 3.0987          & 1.7600          & 0.0666             & 0.3264          & 0.3729          \\
\hline
\textbf{Extra Trees Regressor}           & 1.4312          & 3.1195          & 1.7660          & 0.0602             & 0.3275          & 0.3735          \\
\hline
\textbf{Elastic Net}                     & 1.4703          & 3.2199          & 1.7941          & 0.0308             & 0.3342          & 0.3901          \\
\hline
\textbf{Lasso Regression}                & 1.4709          & 3.2223          & 1.7948          & 0.0300             & 0.3343          & 0.3903          \\
\hline
\textbf{Adaboost Regressor}              & 1.4851          & 3.1357          & 1.7704          & 0.0559             & 0.3216          & \textbf{0.3629}          \\
\hline
\textbf{Dummy Regressor}                 & 1.5051          & 3.3310          & 1.8247          & -0.0021            & 0.3390          & 0.4007          \\
\hline
\end{tabular*}
\label{Tab:t2}
\end{table}

For the sake of comparison, lasso model feature importance for the new target variable has been investigated too. It is observed from Figure 5 that private contribution to pensions by the individual realizes the highest positive coefficient of linear relationship to the self reported risk measure while contribution by employer has the most negative coefficient of linear relationship to the self reported risk measure. The other relevant features with relevant correlation to the target variable are personal income, debt and wealth. Lasso eliminates 50 variables, setting their coefficient to zero in this case, which implies lasso find more variables with significant linear relationship to the self reported risk measure as compared to the average number of safe choices risk measure.

\begin{figure}[!htb]
    \centering
    {\includegraphics[width=0.84\linewidth]{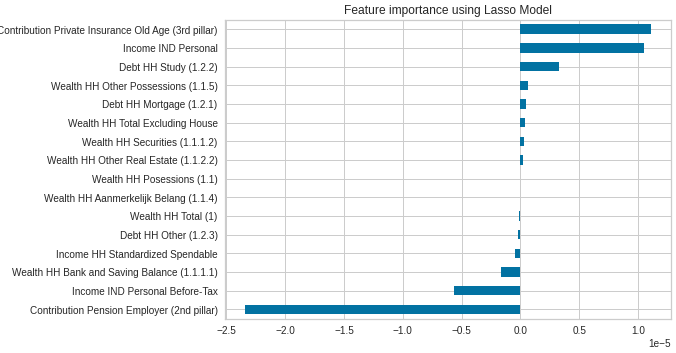} }

    \caption{Lasso regression coefficient of linear relationship to the self reported risk preference measure plot.}\label{fig: figure5}
\end{figure}

Notably, for both target variables, the Lasso regression finds mostly financial variables to be linearly correlated with the target variables, while demographic variables appear to be not as relevant (see Figures 3 and 5) for the estimation of risk preferences. 

For both target variables, ridge regression and LAR models are among the worst performing models indicating that linear models that are susceptible to outliers are not particularly suitable for finding correlations with risk preferences. In general terms, given the results reported in Tables 1 and 2, it seems that models that are susceptible to outliers do perform particularly bad for finding correlations between features and elicited risk preferences.

\section{Discussion}

This study contributes to the investigation of which features are important for individual risk preference estimation and should be considered when doing proper causal analysis. Towards this objective, we recommend the lasso regression model that performs relatively well for both target variables with its ability to eliminate variables unimportant to its linear model in conjunction with the process of robust feature selection carried out in the study provides some insight on which variables are the most relevant. Notably, various models select different features as being relevant with different ordering of levels of importance, which poses a challenge on the selection of most important variables to be considered for modelling.In order to deal with this situation, the top representative models were selected upon which a consensus of features were voted to select the most relevant features. The selected models were lasso regression, gradient boosting machine regressor, and bayesian regression. The first two are the recommended optimal models selected for both target features while also covering closed form solution model and gradient descent based solution. The final model covers a bayesian based solution. Figure 6 provides a box and whisker plot for the distribution of accuracy for each of the selected models using RFECV on the average number of safe choices risk preference measure. 
The topmost performing model accuracy found is about 70\%, which may be considered as relatively low. However, in the case where the prediction precision is not the most relevant factor for deployment such as using the predicted risk measure to recommend appropriate pension savings for an individual, then this moderate accuracy is acceptable for application and deployment.

\begin{figure}[!htb]
    \centering
    {\includegraphics[width=0.64\linewidth]{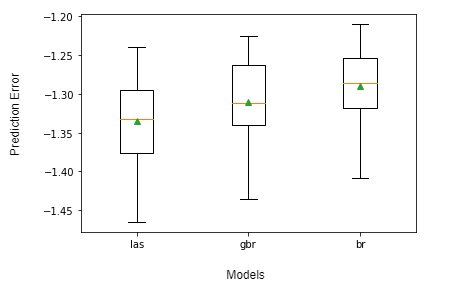} }

    \caption{Box and whisker plot of the performance of lasso, GBR and bayesian regression on the average number of safe choices risk preference measure using recursive feature elimination.}\label{fig: figure6}
\end{figure}

The general trend of performance shows that bayesian regression has the lowest variability of accuracy. The median bar for all the models are also different with the median line of the bayesian regression lying outside the interquartile range box of the lasso regression indicating that each of the models are quite different in their results. Outliers are not overtly present in the results of the three models.

Concerning the self reported risk measure, the box and whisker plot in Figure 7 illustrates that the variability of all the models are quite similar albeit different minima and maxima. The median bars differ reinforcing the difference of results of each model. All three plots also show explicit outlier points which may be due to the fact that self-reported assessments tend to be noisy.

\begin{figure}[!htb]
    \centering
    {\includegraphics[width=0.64\linewidth]{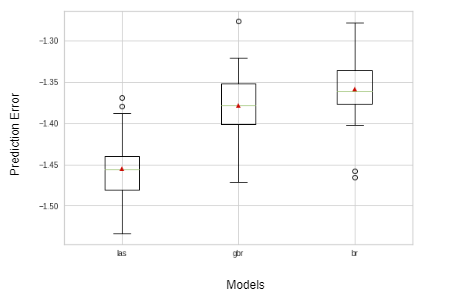} }

    \caption{Box and whisker plot of the performance of lasso, GBR and bayesian regression on the self reported risk preference measure using recursive feature elimination.}\label{fig: figure7}
\end{figure}

The top features that are relevant for the self reported risk measure from the survey using lasso regression are square of the age, pension contribution by employee and employer, private contribution for old age, personal income, household wealth in terms of bank and savings balance and household study debt. For bayesian regression the top relevant features are the square of the age, sex, migration background, individual position in the household, household type, sector of employment (especially for the self-employed), household wealth in percentiles, household spendable income and the presence of children. Finally, for GBR, the top relevant features are the square of the age, sex, sector of employment (especially for the self-employed), individual position in the household, wealth in percentiles together with banking and savings balance, household total debt and household spendable income.

Similarly, the top variables relevant for the average number of safe choices risk preference measure using lasso regression remain the same with the top variables for the self reported risk preference. The only exception is household wealth which is replaced by primary personal income. For bayesian regression the top relevant features are sex, pension fund, individual position in the household, Education level, sector of employment (especially for the self-employed), household wealth in percentiles, household spendable income and the presence of children. This also mostly remains the same with the self reported risk preference with the exception of household type and the square of the age. Lastly, for GBR, the top relevant features are the square of the age, pension fund, pension contribution by employee and employer, personal income, wealth in percentiles together with banking and savings balance and in terms of possession and household spendable income. 

The most consistent feature overall in cross section is the square of the age. This means that this feature is likely to be generally important for risk preference estimation. Other important features are wealth, debt, income, sex, and pension contribution. 
In Figure 2, household wealth in terms of bank and savings balance is positively correlated with the average number of safe choices a participant makes in the MPL survey. This may be due to families willingness to take less risk the higher the buffer required in savings for emergencies. While personal income and household studies debt are strongly negatively correlated with the average number of safe choices. This may be due to the fact that a higher income allows more space and willingness to take risk and invest the extra funds and individuals with student debt are more desperate to earn enough to cover it. On the other hand, in Figure 4, private contribution toward old age pension is strongly correlated positively to the self reported willingness to take risk together with personal income and study debt which preserves the results obtained for the average number of safe choices risk preference measure. Because, an increase in the personal income and debt leads to a  corresponding increase in the likert scale of the willingness to take risk reported by survey participants. Employer contributions towards pension, income before tax and household debt (bank and savings balance) are negatively correlated to the self reported risk measure. 

\subsection{Limitation}
One limitation of this work lies in the fact that the predictive power of the optimal model is still quite low for any sensitive deployment in the applications that directly impacts humans like taking the predicted values as the basis of selecting investment portfolio. However, this is applicable in scenarios where precision is not of uttermost importance, like using the predicted risk preferences as a feature for pension savings recommendation and this study provides a spectrum of baseline models that are applicable for predicting risk preference, which can be further improved upon in future work. Also, the feature importance analysis still requires rigorous causal analysis to make any strong statement on the effect of features on individual risk preference. Whilst, some variables are easier to interpret in a causal way (e.g., sex, age),  others are likely endogenous and not that straight forward (e.g., financial variables).

\section{Conclusion}
The evidence in this article shows that i) prediction of simple elicited risk preference estimated using behavioral economic methods and self reported risk measures by survey participants, have a similar level of prediction power using demographic and financial variables of a representative adult population of the Netherlands. ii) Lasso regression is the recommended optimal model for estimating the average number of safe choices risk measure. This indicates that most of the inputs have a linear relationship with the outcome. The low performance of OLS and ridge regression which are susceptible to outliers suggests that the risk preference measure estimation is highly influenced adversely by the presence of outliers. Gradient boosting machine regressor is the recommended optimal model for prediction of the self reported risk preference measure. We hypothesize that decision errors and noise lead to a higher likelihood of outliers which diminish the performance of models like lasso regression. iii) The topmost prediction accuracy for the prediction of the average number of safe choices risk preference measure is 70.86\% ( MAPE : 0.2904) and that of the self reported risk preference is 63.42\% (MAPE). These findings are baseline and only a first step toward tapping the potential of risk preference prediction. The data are well suited for prediction of other more complex risk preference estimations. This article also touches on which factors are important for risk preference estimation such as age squared, wealth, income, debt etc. However, this work does not measure the effect of the variables, it just provides suggestions on interesting variables that are worth investigating their effect on risk preference.

\clearpage
\begin{backmatter}
\section*{Declarations}

\section*{Availability of data and materials}
The administrative input dataset from Statistics Netherlands analysed during the current study are not publicly available due privacy concerns of the official statistics management organization policy for the Netherlands but are available from the corresponding author in collaboration with Statistics Netherlands on reasonable request and agreement.

\section*{Competing interests}
  The authors declare that they have no competing interests.

\section*{Funding}
Financial support from Netspar through the grant “Understanding and improving pension savings by combining incentivized experiments, surveys, and administrative big data - A general employed population sample with a focus on the self-employed” is gratefully acknowledged. Ethical approval was obtained from the Ethical Review Committee Inner City faculties (ERCIC): reference ERCIC 182 27 02 2020.

\section*{Author's contributions}
    O.A. carried out all the analysis and wrote the first draft of the paper. A.R. edited and revised the draft and supervised the work. L.I. and M.D. participated in the revision and supervision of the work.

\section*{Acknowledgements}
  This research has been funded by Netspar (Network for Studies on Pensions, Aging and Retirement) 2019 theme grant. I would like to express my special thanks
of gratitude to my advisors and collaborators Michel Dumontier, Arno Riedl, Lianne Ippel, Chang Sun, Leto Peel, Marco Puts and Paul Bokern.

\renewcommand{\nomname}{List of Abbreviations}
\nomenclature{\(ML\)}{Machine Learning}
\nomenclature{\(NN\)}{Neural Networks}
\nomenclature{\(OLS\)}{Ordinary Least Square}
\nomenclature{\(LARS\)}{Least Angle Regression}
\nomenclature{\(OMP\)}{Orthogonal Matching Pursuit}
\nomenclature{\(SVM\)}{Support Vector Machine}
\nomenclature{\(XGBoost\)}{Extreme Gradient Boosting}
\nomenclature{\(LightGBM\)}{Light Gradient Boosting Machine}
\nomenclature{\(GBR\)}{Gradient Boosting}
\nomenclature{\(RF\)}{Random Forest}
\nomenclature{\(adaBoost\)}{Adaptive Boosting}
\nomenclature{\(CatBoost\)}{Category Boosting}
\nomenclature{\(CBS\)}{Statistics Netherlands}
\nomenclature{\(MPL\)}{Multiple Price List}
\nomenclature{\(RFECV\)}{Recursive feature elimination with cross validation}
\nomenclature{\(MAE\)}{Mean Absolute Error}
\nomenclature{\(MAPE\)}{Mean absolute Percentage Error}
\nomenclature{\(MSE\)}{Mean Squared Error}
\nomenclature{\(RMSE\)}{Root Mean Squared Error}
\nomenclature{\(RMSLE\)}{Root Mean Squared Log Error}
\nomenclature{\(R2\)}{R-Squared}
\printnomenclature

\clearpage


\bibliographystyle{bmc-mathphys} 
\bibliography{bmc_article}      

\clearpage
\section*{Appendix}
\subsection*{A1. Variables description}
{\tiny
\begin{longtable}{p{5cm}cc}
\caption{Data dictionary of register data from statistics Netherlands with missing data percentage.}    \\
\textbf{Data Variable (Statistics Netherlands Code)}  &   \textbf{Description}     & \textbf{Missing data (\%)}    \\
\hline \hline
LFTENQ2                         & Age                                                    & 0.14           \\
MIGRATIEACHTERGROND             & Migration Background                                   & 0.14            \\
TYPHH2                          & House Hold Type                                        & 0.14           \\
PLHH2                           & House Hold Position of Individual                      & 0.14    \\
AANTALPERSHH2                   & House Hold Size                                        & 0.14           \\
AANTALKINDHH2                   & House Hold Amount of Children                          & 0.14           \\
SBI2                            & Sector (SBI) Employee                                  & 0.14          \\
SMODELRAMINGPF2                 & Pension Fund                                           & 0.14          \\
SMODELRAMINGPENSIOENPREMIEWG2   & Pension Contribution Employer                          & 36.27         \\
SMODELRAMINGPENSIOENPREMIEWN2   & Pension Contribution Employee                          & 36.27         \\
INPSECJ2019                     & Occupation (Social-Economic Category)                  & 0.00          \\
OCCUPATION2019                  & Occupation 4 Categories                                & 0.00          \\
INPZELFSTANDIGEPL12019          & Occupation Publication Classification Self-Employed    & 0.00          \\
INPTYPZLF2019                   & Occupation Type of Self-Employed                       & 0.00          \\
SBISELFEMPLOYED2019             & Sector (SBI) All Types Self-Employed                   & 0.00          \\
INPPN700PEN2019                 & Contribution Pension Employee (2nd pillar)             & 0.00          \\
INPPG710PEN2019                 & Contribution Pension Employer (2nd pillar)             & 0.00          \\
INPPH770OUP2019                 & Contribution Private Insurance Old Age (3rd pillar)    & 0.00          \\
INPPH570ZWP2019                 & Contribution Private Insurance Incapacitation          & 0.00          \\
INPPINK2019                     & Income Individual Personal Y/N                         & 0.00          \\
INPPERSPRIM2019                 & Income Individual Personal Primary                     & 0.00         \\
INPPERSINK2019                  & Income Individual Personal                             & 0.14          \\
INPPERSBRUT2019                 & Income Individual Personal Before-Tax                  & 0.14          \\
TYPHH2019                       & House Hold Type                                        & 0.14          \\
PLHH2019                        & House Hold Position of Individual                      & 0.14   \\
AANTALPERSHH2019                & House Hold Size                                        & 0.14          \\
AANTALKINDHH2019                & House Hold Amount of Children                          & 0.14          \\
GBABURGSTNWKLASSE42019          & Marital Status 4 Categories                            & 0.14          \\
VEHP100HVERM2019                & Wealth House Hold Percentiles                          & 0.14          \\
VEHP100HVERMKL12019             & Wealth House Hold Deciles                              & 0.14          \\
VEHWVEREXEWH2019                & Wealth House Hold Total Excluding House                & 0.14          \\
VEHW1000VERH2019                & Wealth House Hold Total (1)                            & 0.14          \\
VEHW1100BEZH2019                & Wealth House Hold Posessions (1.1)                     & 0.14          \\
VEHW1110FINH2019                & Wealth House Hold Financial Possessions (1.1.1)        & 0.14          \\
VEHW1111BANH2019                & Wealth House Hold Bank and Saving Balance (1.1.1.1)    & 0.14          \\
VEHW1112EFFH2019                & Wealth House Hold Securities (1.1.1.2)                 & 0.14          \\
SECURITIESPERC2019              & Securities \% of liquid wealth                         & 0.14          \\
VEHW1120ONRH2019                & Wealth House Hold Real Estate (1.1.2)                  & 0.14          \\
VEHW1121WONH2019                & Wealth House Hold House (1.1.2.1)                      & 0.14          \\
VEHW1122OGOH2019                & Wealth House Hold Other Real Estate (1.1.2.2)          & 0.14          \\
VEHW1130ONDH2019                & Wealth House Hold Entrepreneurial Capacity (1.1.3)     & 0.14          \\
VEHW1140ABEH2019                & Wealth House Hold Aanmerkelijk Belang (1.1.4)          & 0.14          \\
VEHW1150OVEH2019                & Wealth House Hold Other Possessions (1.1.5)            & 0.14          \\
VEHW1200STOH2019                & Debt House Hold total (1.2)                            & 0.14          \\
VEHW1210SHYH2019                & Debt House Hold Mortgage (1.2.1)                       & 0.14          \\
VEHW1220SSTH2019                & Debt House Hold Study (1.2.2)                          & 0.14          \\
VEHW1230SOVH2019                & Debt House Hold Other (1.2.3)                          & 0.14          \\
INHEHALGR2019                   & House Hold Homeowner                                   & 0.14          \\
INHP100HGEST2019                & Income House Hold Standardized Spendable Percentiles   & 0.14          \\
INHP100HGESTKL12019             & Income House Hold Standardized Spendable Deciles       & 0.14          \\
INHGESTINKH2019                 & Income House Hold Standardized Spendable               & 0.14          \\
OPLNIVSOI2016AGG1HBMETNIRWO2019 & Education Level 3 Categories                           & 0.14          \\
INPPOSHHK2019                   & Position in household towards main breadwinner         & 0.14          \\
INHBBIHJ2019                    & Main source of household income                        & 0.14          \\
NRCHILDREN2019                  & Number of Children                                     & 0.14          \\
NRCHILDRENSAMEADRS2019          & Number of Children at Same Address as Individual       & 31.90\\
NRCHILDRENCOPARENTADRS2019      & Number of Children at Same Address as Co-Parent        & 31.90 \\
DEADBORN2019                    & Indicates Whether Child(ren) Was(Were) Deadborn        & 31.90 \\
NRCOPARENTS2019                 & Amount of People the Individual had Children With  & 31.90\\
LASTCOPARENTSAMEADRS2019        & Inidicates Whether Last Co-Parent at Same Address as Individual  & 31.90\\
Age\_squared                    & Squared value of the age                               & 0.14          \\
GBAGESLACHT                     & Sex                                                    & 0.00          \\
MAINBREADWINNER2019             & Main Breadwinner                                       & 0.00          \\
SECURITIESBIN2019               & Indicates whether household has securities             & 0.00   \\
HOMEOWNER2019\_                 & Home owner                                             & 0.00          \\
CHILDBIN2019                    & Number of children born                                & 0.00         
\label{Tab:t5}
\end{longtable}
}

\subsection*{A2. Full set of models trained and metrics tables}
\begin{table}[!htb]
\caption{Full list of machine learning models for prediction of the average number of safe choices risk preference measure and the corresponding regression metric values.}
\centering
\resizebox{\textwidth}{!}{
\begin{tabular}{||c c c c c c c||} 
\hline \hline
\textbf{Model}                           & \textbf{MAE}    & \textbf{MSE}    & \textbf{RMSE}   & \textbf{R-Squared} & \textbf{RMSLE}  & \textbf{MAPE}   \\ 
\hline 
\textbf{Orthogonal Matching Pursuit}     & \textbf{1.3273} & \textbf{2.7990} & \textbf{1.6713} & \textbf{0.0140}    & \textbf{0.2786} & \textbf{0.2904} \\
\hline
\textbf{Elastic Net}                     & 1.3299          & 2.8111          & 1.6748          & 0.0097             & 0.2793          & 0.2909          \\
\hline
\textbf{Lasso Regression}                & 1.3306          & 2.8166          & 1.6765          & \textbf{0.0078}    & 0.2794          & \textbf{0.2908} \\
\hline
\textbf{Bayesian Ridge}                  & 1.3312          & 2.8198          & 1.6775          & 0.0067             & 0.2794          & 0.2908          \\
\hline
\textbf{Adaboost Regressor}              & 1.3418          & 2.8399          & 1.6833          & -0.0001            & 0.2813          & 0.2965          \\
\hline
\textbf{Dummy Regressor}                 & 1.3359          & 2.8422          & 1.6841          & -0.0011            & 0.2802          & 0.2916          \\
\hline
\textbf{Lasso Least Angle Regression}    & 1.3359          & 2.8422          & 1.6841          & -0.0011            & 0.2802          & 0.2916          \\
\hline
\textbf{Random Forest Regressor}         & 1.3406          & 2.8567          & 1.6885          & -0.0068            & 0.2809          & 0.2919          \\
\hline
\textbf{Gradient Boosting Regressor}     & 1.3457          & 2.8643          & 1.6903          & -0.0082            & 0.2813          & 0.2939          \\
\hline
\textbf{Catboost Regressor}              & 1.3660          & 2.9519          & 1.7164          & -0.0402            & 0.2845          & 0.2958          \\
\hline
\textbf{Extra Trees Regressor}           & 1.3734          & 2.9760          & 1.7233          & -0.0488            & 0.2855          & 0.2985          \\
\hline
\textbf{Light Gradient Boosting Machine} & 1.3800          & 3.0155          & 1.7352          & -0.0637            & 0.2870          & 0.2996          \\
\hline
\textbf{KNN Regressor}                   & 1.4405          & 3.2923          & 1.8130          & -0.1622            & 0.2973          & 0.3134          \\
\hline
\textbf{Decision Tree}                   & 1.9044          & 5.6967          & 2.3840          & -1.0118            & 0.4033          & 0.3945          \\
\hline
\textbf{Huber Regressor}                 & 2.4010          & 11.7542         & 3.3986          & -3.1356            & 0.4948          & 0.4519          \\
\hline
\textbf{Linear Regression}               & 1.4927          & 34.7611         & 3.3179          & -12.2377           & 0.2926          & 0.3136          \\
\hline
\textbf{Passive Aggressive Regressor}    & 9.9817          & 885.4038        & 22.2222         & -307.5104          & 0.9122          & 1.9582          \\
\hline
\textbf{Ridge Regression}                & 3.8896          & 9451.7096       & 33.1493         & -3622.3047         & 0.3182          & 0.5478         \\
\hline
\end{tabular}}
\label{Tab:t3}
\end{table}

\begin{table}[!htb]
\caption{Full list of machine learning models for prediction of self reported risk preference measure and the corresponding regression metric values.}
\centering
\resizebox{\textwidth}{!}{
\begin{tabular}{||c c c c c c c||} 
\hline \hline
\textbf{Model}                           & \textbf{MAE}    & \textbf{MSE}    & \textbf{RMSE}   & \textbf{R-Squared} & \textbf{RMSLE}  & \textbf{MAPE}   \\
\hline
\textbf{Orthogonal Matching Pursuit}     & \textbf{1.3846} & \textbf{2.9033} & \textbf{1.7037} & \textbf{0.1258}    & \textbf{0.3199} & 0.3658          \\
\hline
\textbf{Elastic Net}                     & 1.4703          & 3.2199          & 1.7941          & 0.0308             & 0.3342          & 0.3901          \\
\hline
\textbf{Lasso Regression}                & 1.4709          & 3.2223          & 1.7948          & 0.0300             & 0.3343          & 0.3903          \\
\hline
\textbf{Bayesian Ridge}                  & 1.3856          & 3.1209          & 1.7597          & 0.0617             & 0.3205          & 0.3647          \\
\hline
\textbf{Adaboost Regressor}              & 1.4851          & 3.1357          & 1.7704          & 0.0559             & 0.3216          & 0.3629          \\
\hline
\textbf{Dummy Regressor}                 & 1.5051          & 3.3310          & 1.8247          & -0.0021            & 0.3390          & 0.4007          \\
\hline
\textbf{Lasso Least Angle Regression}    & 1.5051          & 3.3310          & 1.8247          & -0.0021            & 0.3390          & 0.4007          \\
\hline
\textbf{Random Forest Regressor}         & 1.4044          & 2.9644          & 1.7215          & 0.1074             & 0.3212          & 0.3680          \\
\hline
\textbf{Gradient Boosting Regressor}     & 1.3847          & 2.9183          & 1.7081          & 0.1213             & 0.3201          & \textbf{0.3656} \\
\hline
\textbf{Catboost Regressor}              & 1.4234          & 3.0767          & 1.7537          & 0.0739             & 0.3248          & 0.3700          \\
\hline
\textbf{Extra Trees Regressor}           & 1.4312          & 3.1195          & 1.7660          & 0.0602             & 0.3275          & 0.3735          \\
\hline
\textbf{Light Gradient Boosting Machine} & 1.4259          & 3.0987          & 1.7600          & 0.0666             & 0.3264          & 0.3729          \\
\hline
\textbf{KNN Regressor}                   & 1.5975          & 3.8369          & 1.9580          & -0.1550            & 0.3543          & 0.4063          \\
\hline
\textbf{Decision Tree}                   & 1.9625          & 6.1450          & 2.4475          & -0.8527            & 0.4685          & 0.4796          \\
\hline
\textbf{Huber Regressor}                 & 2.2803          & 9.8726          & 3.1179          & -1.9675            & 0.4900          & 0.4900          \\
\hline
\textbf{Linear Regression}               & 2.7323          & 1574.0018       & 14.6457         & -481.7652          & 0.4194          & 0.7557          \\
\hline
\textbf{Passive Aggressive Regressor}    & 7.3215          & 389.1269        & 17.4595         & -117.0008          & 0.7994          & 1.4651          \\
\hline
\textbf{Ridge Regression}                & 16.5485         & 3.4373 * 105    & 1888.5328       & -1.0543 * 105      & 0.3840          & 5.3980          \\
\hline
\end{tabular}}
\label{Tab:t4}
\end{table}

\subsection*{A3. Multiple Price List}
\subsubsection*{MPL1}
\begin{table}[!hp]
\caption{Multiple price list variation 1.}
\centering
\resizebox{\textwidth}{!}{
\begin{tabular}{p{1cm}p{1cm}p{1cm}p{1cm}p{1cm}p{1cm}p{1cm}p{1cm}p{1cm}p{1cm}p{1cm}}
\textbf{Lottery} & \multicolumn{5}{c}{OPTION A}                               & \multicolumn{5}{c}{OPTION B}                               \\
\hline \hline
                 & Probablity & Amount & Probablity & Amount & Expected value & Probablity & Amount & Probablity & Amount & Expected value \\
\hline \hline
\#1              & 0.1        & €80    & 0.9        & €64    & €66            & 0.1        & €154   & 0.9        & €4     & €19            \\
\#2              & 0.2        & €80    & 0.8        & €64    & €67            & 0.2        & €154   & 0.8        & €4     & €34            \\
\#3              & 0.3        & €80    & 0.7        & €64    & €69            & 0.3        & €154   & 0.7        & €4     & €49            \\
\#4              & 0.4        & €80    & 0.6        & €64    & €70            & 0.4        & €154   & 0.6        & €4     & €64            \\
\#5              & 0.5        & €80    & 0.5        & €64    & €72            & 0.5        & €154   & 0.5        & €4     & €79            \\
\#6              & 0.6        & €80    & 0.4        & €64    & €74            & 0.6        & €154   & 0.4        & €4     & €94            \\
\#7              & 0.7        & €80    & 0.3        & €64    & €75            & 0.7        & €154   & 0.3        & €4     & €109           \\
\#8              & 0.8        & €80    & 0.2        & €64    & €77            & 0.8        & €154   & 0.2        & €4     & €124           \\
\#9              & 0.9        & €80    & 0.1        & €64    & €78            & 0.9        & €154   & 0.1        & €4     & €139           \\
\#10             & 1          & €80    & 0          & €64    & €80            & 1          & €154   & 0          & €4     & €154          
\end{tabular}}
\label{Tab:t6}
\end{table}

\subsubsection*{MPL2}
\begin{table}[!htb]
\caption{Multiple price list variation 2.}
\centering
\resizebox{\textwidth}{!}{
\begin{tabular}{p{1cm}p{1cm}p{1cm}p{1cm}p{1cm}p{1cm}p{1cm}p{1cm}p{1cm}p{1cm}p{1cm}}
\textbf{Lottery} & \multicolumn{5}{c}{OPTION A}                               & \multicolumn{5}{c}{OPTION B}                               \\
\hline \hline
                 & Probablity & Amount & Probablity & Amount & Expected value & Probablity & Amount & Probablity & Amount & Expected value \\
\hline \hline
\#1              & 0.1        & €99    & 0.9        & €41    & €47            & 0.1        & €134   & 0.9        & €19    & €31            \\
\#2              & 0.2        & €99    & 0.8        & €41    & €53            & 0.2        & €134   & 0.8        & €19    & €42            \\
\#3              & 0.3        & €99    & 0.7        & €41    & €58            & 0.3        & €134   & 0.7        & €19    & €54            \\
\#4              & 0.4        & €99    & 0.6        & €41    & €64            & 0.4        & €134   & 0.6        & €19    & €65            \\
\#5              & 0.5        & €99    & 0.5        & €41    & €70            & 0.5        & €134   & 0.5        & €19    & €77            \\
\#6              & 0.6        & €99    & 0.4        & €41    & €76            & 0.6        & €134   & 0.4        & €19    & €88            \\
\#7              & 0.7        & €99    & 0.3        & €41    & €82            & 0.7        & €134   & 0.3        & €19    & €100           \\
\#8              & 0.8        & €99    & 0.2        & €41    & €87            & 0.8        & €134   & 0.2        & €19    & €111           \\
\#9              & 0.9        & €99    & 0.1        & €41    & €93            & 0.9        & €134   & 0.1        & €19    & €123           \\
\#10             & 1          & €99    & 0          & €41    & €99            & 1          & €134   & 0          & €19    & €134          
\end{tabular}}
\label{Tab:t7}
\end{table}

\subsubsection*{MPL3}
\begin{table}[!htb]
\caption{Multiple price list variation 3.}
\centering
\resizebox{\textwidth}{!}{
\begin{tabular}{p{1cm}p{1cm}p{1cm}p{1cm}p{1cm}p{1cm}p{1cm}p{1cm}p{1cm}p{1cm}p{1cm}}
\textbf{Lottery} & \multicolumn{5}{c}{OPTION A}                               & \multicolumn{5}{c}{OPTION B}                               \\
\hline \hline
                 & Probablity & Amount & Probablity & Amount & Expected value & Probablity & Amount & Probablity & Amount & Expected value \\
\hline \hline
\#1              & 1          & €52    &            &        & €52            & 0.5        & €30    & 0.5        & €130   & €80            \\
\#2              & 1          & €57    &            &        & €57            & 0.5        & €30    & 0.5        & €130   & €80            \\
\#3              & 1          & €63    &            &        & €63            & 0.5        & €30    & 0.5        & €130   & €80            \\
\#4              & 1          & €68    &            &        & €68            & 0.5        & €30    & 0.5        & €130   & €80            \\
\#5              & 1          & €73    &            &        & €73            & 0.5        & €30    & 0.5        & €130   & €80            \\
\#6              & 1          & €78    &            &        & €78            & 0.5        & €30    & 0.5        & €130   & €80            \\
\#7              & 1          & €82    &            &        & €82            & 0.5        & €30    & 0.5        & €130   & €80            \\
\#8              & 1          & €88    &            &        & €88            & 0.5        & €30    & 0.5        & €130   & €80            \\
\#9              & 1          & €94    &            &        & €94            & 0.5        & €30    & 0.5        & €130   & €80            \\
\#10             & 1          & €101   &            &        & €101           & 0.5        & €30    & 0.5        & €130   & €80           
\end{tabular}}
\label{Tab:t8}
\end{table}

\subsubsection*{MPL4}
\begin{table}[!htb]
\caption{Multiple price list variation 4.}
\centering
\resizebox{\textwidth}{!}{
\begin{tabular}{p{1cm}p{1cm}p{1cm}p{1cm}p{1cm}p{1cm}p{1cm}p{1cm}p{1cm}p{1cm}p{1cm}}
\textbf{Lottery} & \multicolumn{5}{c}{OPTION A}                               & \multicolumn{5}{c}{OPTION B}                               \\
\hline \hline
                 & Probablity & Amount & Probablity & Amount & Expected value & Probablity & Amount & Probablity & Amount & Expected value \\
\hline \hline
\#1              & 1          & €39    &            &        & €39            & 0.33       & €20    & 0.67       & €110   & €80            \\
\#2              & 1          & €46    &            &        & €46            & 0.33       & €20    & 0.67       & €110   & €80            \\
\#3              & 1          & €56    &            &        & €56            & 0.33       & €20    & 0.67       & €110   & €80            \\
\#4              & 1          & €64    &            &        & €64            & 0.33       & €20    & 0,67       & €110   & €80            \\
\#5              & 1          & €70    &            &        & €70            & 0.33       & €20    & 0,67       & €110   & €80            \\
\#6              & 1          & €75    &            &        & €75            & 0.33       & €20    & 0,67       & €110   & €80            \\
\#7              & 1          & €79    &            &        & €79            & 0.33       & €20    & 0,67       & €110   & €80            \\
\#8              & 1          & €84    &            &        & €84            & 0.33       & €20    & 0,67       & €110   & €80            \\
\#9              & 1          & €88    &            &        & €88            & 0.33       & €20    & 0,67       & €110   & €80            \\
\#10             & 1          & €93    &            &        & €93            & 0.33       & €20    & 0,67       & €110   & €80           
\end{tabular}}
\label{Tab:t9}
\end{table}

\subsubsection*{MPL5}
\begin{table}[!htb]
\caption{Multiple price list variation 5.}
\centering
\resizebox{\textwidth}{!}{
\begin{tabular}{p{1cm}p{1cm}p{1cm}p{1cm}p{1cm}p{1cm}p{1cm}p{1cm}p{1cm}p{1cm}p{1cm}}
\textbf{Lottery} & \multicolumn{5}{c}{OPTION A}                               & \multicolumn{5}{c}{OPTION B}                               \\
\hline \hline
                 & Probablity & Amount & Probablity & Amount & Expected value & Probablity & Amount & Probablity & Amount & Expected value \\
\hline \hline
\#1              & 0.5        & €90    & 0.5        & €70    & €80            & 0.5        & €103   & 0.5        & €35    & €69            \\
\#2              & 0.5        & €90    & 0.5        & €70    & €80            & 0.5        & €109   & 0.5        & €35    & €72            \\
\#3              & 0.5        & €90    & 0.5        & €70    & €80            & 0.5        & €115   & 0.5        & €35    & €75            \\
\#4              & 0.5        & €90    & 0.5        & €70    & €80            & 0.5        & €122   & 0.5        & €35    & €79            \\
\#5              & 0.5        & €90    & 0.5        & €70    & €80            & 0.5        & €128   & 0.5        & €35    & €82            \\
\#6              & 0.5        & €90    & 0.5        & €70    & €80            & 0.5        & €131   & 0.5        & €35    & €83            \\
\#7              & 0.5        & €90    & 0.5        & €70    & €80            & 0.5        & €138   & 0.5        & €35    & €87            \\
\#8              & 0.5        & €90    & 0.5        & €70    & €80            & 0.5        & €153   & 0.5        & €35    & €94            \\
\#9              & 0.5        & €90    & 0.5        & €70    & €80            & 0.5        & €170   & 0.5        & €35    & €103           \\
\#10             & 0.5        & €90    & 0.5        & €70    & €80            & 0.5        & €186   & 0.5        & €35    & €111          
\end{tabular}}
\label{Tab:t10}
\end{table}

\subsection*{A4. Survey General risk question}
\begin{table}[!htb]
\caption{Sample of general risk question from survey}
\resizebox{\textwidth}{!}{
\begin{tabular}{ll}
\hline \hline
\textbf{Risk}                           & Scale: 0 "not at all willing to take risks" – 10 "very willing to take risks"                                                                                               \\ \hline \hline \cline{1-1}
\multicolumn{1}{|l|}{General}           & \begin{tabular}[c]{@{}l@{}}Can you tell me to what extent you are, in general, willing or unwilling\\ are to take risks?\end{tabular}                                       \\ \cline{1-1}
\multicolumn{1}{|l|}{Domain-Specific}   & \begin{tabular}[c]{@{}l@{}}People can behave differently in different situations. How do you\\ assess your willingness to take risks in the following matters:\end{tabular} \\ \cline{1-1}
\multicolumn{1}{|l|}{Occupation}        & … in your career choice?                                                                                                                                                    \\ \cline{1-1}
\multicolumn{1}{|l|}{Health}            & … in your health? {[}with option “not applicable”{]}                                                                                                                        \\ \cline{1-1}
\multicolumn{1}{|l|}{Personal Finances} & … in your personal financial affairs?                                                                                                                                       \\ \cline{1-1}
\multicolumn{1}{|l|}{Job finances}      & … in your work-related financial matters?                                                                                                                                   \\ \cline{1-1}
\end{tabular}}
\label{Tab:t11}
\end{table}




\end{backmatter}
\end{document}